\documentclass[twocolumn,amssymb,nobibnotes,aps,prb]{revtex4-2}

\usepackage[dvipsnames]{xcolor}

\usepackage{graphicx}
\usepackage{natbib}
\usepackage{url}
\usepackage{textcase}
\usepackage{bm}
\usepackage{xcolor}
\usepackage{braket}
\usepackage{amsmath}
\usepackage{amssymb}
\usepackage[breaklinks]{hyperref}
\hypersetup{
  colorlinks = true,
  linkcolor = blue,
  citecolor = blue
}

\newcommand{\w}{\omega}
\newcommand{\vk}{\mathbf{k}}
\newcommand{\vecr}{\mathbf{r}}
\newcommand{\vq}{\mathbf{q}}
\newcommand{\vG}{\mathbf{G}}

\begin{document}

\title{Magnons in chromium trihalides from \emph{ab initio} Bethe-Salpeter equation}


\author{Ali Esquembre-Ku\v{c}ukali\'c$^1$}
\author{Khoa B. Le$^2$}
\author{Alberto Garc\'ia-Crist\'obal$^1$}
\author{Marco Bernardi$^2$}
\author{Davide Sangalli$^3$}
\author{Alejandro Molina-S\'anchez$^1$}

\address{$^1$Institute of Materials Science (ICMUV), University of Valencia,  Catedr\'{a}tico Beltr\'{a}n 2,  E-46980,  Valencia,  Spain}
\address{$^2$Department of Applied Physics and Materials Science, and Department of Physics, California Institute of Technology, Pasadena, California 91125, USA,}
\address{$^3$Istituto di Struttura della Materia-CNR (ISM-CNR), Area della Ricerca di Roma 1, Monterotondo Scalo, Italy}

\begin{abstract}
    
Chromium trihalides (CrX$_3$, with $\rm{X=I,Br,Cl}$) are layered ferromagnetic materials with rich physics and possible applications. Their structure consists of magnetic Cr atoms positioned between two layers of halide atoms. The choice of halide results in distinct magnetic properties, but their effect on spin-wave (magnon) excitations is not fully understood. Here we present first-principles calculations of magnon dispersions and wave functions for monolayer Cr trihalides using the finite-momentum Bethe-Salpeter equation (BSE) to describe collective spin-flip excitations. 
We study the dependence of magnon dispersions on the halide species and resolve the small topological gap at the Dirac point in the magnon spectrum by including spin-orbit coupling. Analysis of magnon wave functions reveals that magnons are made up of electronic transitions with a wider energy range than excitons in CrX$_3$ monolayers, providing insight into magnon states in real and reciprocal space. We discuss Heisenberg exchange parameters extracted from the BSE and discuss the convergence of BSE magnon calculations. Our work advances the quantitative modeling of magnons in two-dimensional materials, providing the starting point for studying magnon interactions in a first-principles BSE framework.
\end{abstract}

\maketitle

\section*{Introduction}
\vspace{-10pt}
The discovery of magnetic order in two-dimensional (2D) materials has opened new directions in the exploration of magnetism in atomically thin structures \cite{Huang2017,Gibertini2019, Li2022}. 
The family of 2D magnetic materials is growing rapidly~\cite{Burch2018, Wang2022}: notable systems studied so far include the out-of-plane ferromagnets CrI$_3$, CrBr$_3$ and Cr$_2$Ge$_2$Te$_6$~\cite{Huang2017, Zhang2019, Gong2017}, the in-plane ferromagnet CrCl$_3$ ~\cite{BedoyaPinto2021}, the layered transition-metal thiophosphate antiferromagnets MPS$_3$ ($\rm{M=Mn, Fe, Ni}$)\cite{MPS}, and systems such as CrSBr with ferromagnetic monolayer and antiferromagnetic multilayers~\cite{Lee2021}. 
\\
\indent
Layered magnetic materials can be stacked in van der Waals heterostructures, providing a material platform for novel device physics in the ultrathin limit with applications to spintronics, spin valves and other spin-based quantum technologies~\cite{Cardoso2018, Soriano2020}. 
Magnetic phenomena such as proximity effects \cite{Fu2020}, chiral magnetic structures \cite{RoldanMolina2016} and magnetic topological phases \cite{Wang2018,Chen2018} have also been studied in connection with novel devices and applications, including random access memory devices and quantum computing applications~\cite{Hirohata2020,Bhatti2017,Yuan2022}.
\\
\indent
In this rapidly evolving landscape of spin-based technologies, theory and simulations play an important role in the prediction of magnetic ground-state properties and spin-wave excitations (magnons), and in the interpretation of experimental data. In particular, magnons have been studied for decades \cite{Bloch1930}, but only recently they have been analyzed in detail in bulk and 2D magnetic materials relevant for spintronics~\cite{Chumak2015}. Linear spin-wave theory (LSWT) calculations of magnon spectra have become widely used following the development of open-source codes, such as \textsc{SpinW}~\cite{Toth2015}, based on model spin Hamiltonians. 
In addition, codes such as \textsc{TB2J}~\cite{He2021} can compute the exchange parameters for the Heisenberg model using density functional theory (DFT). Combinations of DFT and model Hamiltonians have recently been used to describe the magnetic anisotropy in CrI$_3$ \cite{Lado2017} and to characterize the behavior of spin waves in CrSBr under strain \cite{Esteras2022}, while DFT combined with atomic orbital projections \cite{Costa2020} or Wannier Hamiltonians \cite{Ferron2015} has provided accurate tight-binding descriptions of CrI$_3$ topological magnons and enabled simulations of magnetic impurities. Finally, combining atomistic spin models and micromagnetic calculations based on the Landau-Lifshitz-Gilbert equation is proving useful to study the rich physics of magnetic nanostructured materials \cite{Evans2014}. 
\\
\indent
However, \textit{ab initio} calculations of magnons in real materials are still in their early days. Existing calculations are primarily based on time-dependent DFT (TD-DFT) \cite{Gorni2018, TancogneDejean2020, Skovhus2022, Gorni2022}, with examples including simulations of ultrafast demagnetization \cite{Krieger2015,Simoni2017,Krieger2017} and calculations of magnon-phonon coupling in CrI$_3$ monolayer \cite{Delugas2023} and the magnon Dirac gap in CrI$_3$~\cite{Gorni2023}. In the framework of many-body perturbation theory, Green's function methods are widely used to model optical excitations~\cite{Onida2002}, but calculations
of magnetic excitations are still scarce~\cite{Kotani2008,Muller2016,Olsen2021}. 
\\
\indent
Here, we employ a many-body formalism to study magnons in Cr trihalides monolayers (CrX$_3$, with $\rm{X=I,Br,Cl}$) using the Bethe-Salpeter equation (BSE). 
Magnons are treated as collective electron-hole excitations, similar to excitons, and the magnon dispersions and wave functions are computed in the transition basis from the BSE Hamiltonian. 
This allows us to analyze magnon features beyond the dispersion, including the magnon wave function in real and reciprocal space, and to compare excitons and magnons in CrX$_3$ monolayers using optical and spin response functions. 
In addition, we include spin-orbit coupling (SOC) using a fully spinorial BSE formulation~\cite{Marsili2021}, allowing the study of the effects of SOC on the Dirac gap. The effect of the halide atoms in the magnon dispersions is also analyzed. Finally, we conclude by discussing the convergence of our results from the BSE. Our detailed magnon calculations in Cr trihalides show that the BSE is a valuable tool for studying magnons and set the stage for the further theoretical development, that we have applied in the study of magnon-phonon interactions in our companion paper \cite{companion}. 

\section{Magnons in the Bethe-Salpeter equation formalism}  \label{sec:theory}
\vspace{-10pt}
We briefly review magnons in the BSE framework and calculations of optical and spin response functions. 
The BSE describes excitons as a coherent superposition of electron-hole excitations, with spin conservation resulting from selection rules for light-matter interaction. Consequently, optical excitations in 2D magnetic semiconductors, such as Cr trihalides, are dominated by singlet excitons, where the electron and hole possess the same spin \cite{Wu2019,Molina2020}. 
In contrast, in the BSE formalism, magnons are viewed as coherent superpositions of electron-hole excitations with opposite electron and hole spins. Early attempts to compute magnons with the BSE~\cite{Aryasetiawan1999,Karlsson2000} were followed by works focusing on simple magnetic materials such as iron, cobalt, and nickel~\cite{Sastoglu2010,Muller2016}. 
Only very recently, the BSE has been employed to study magnons in a 2D material, CrI$_3$~\cite{Olsen2021}. 
\\
\indent
In the BSE, one diagonalizes an effective two-particle Hamiltonian consisting of a non-interacting electron-hole term plus a kernel containing direct and exchange electron-hole interactions~\cite{Onida2002}. Solving the BSE for excitons and magnons requires computing the electronic band structure and wave functions. 
Here, we employ band structures extracted from DFT and obtain quasiparticle corrections using the G$_0$W$_0$ approximation \cite{Marini2009} as a scissor operator. The spinorial BSE used in this work, with or without the inclusion of SOC, has recently been implemented in the \textsc{Yambo} code \cite{Marini2009, Sangalli2019}, which had been used to study excitons in magnetic and nonmagnetic systems \cite{DiSabatino2023,Sayers2023,Marsili2021, Galvani2016}. 
\\
\indent
To write the BSE Hamiltonian using a compact notation, we denote electron-hole transitions using ${\{I,\vq\}=\{v\vk-\vq \rightarrow c\vk\}}$, where $v$ and $c$ label valence and conduction bands, $\vk$ is the electron crystal momentum, and $\vq$ is the transferred momentum in the electronic transition. The band indices include spin when SOC is neglected or label full spinors when SOC is included. The corresponding transition energies for non-interacting electron-hole pairs are ${\Delta \epsilon_I(\vq) = \epsilon_{c\vk} - \epsilon_{v\vk-\vq}}$ and the associated occupation factors are ${\Delta f_I(\vq)=f_{v\vk-\vq} - f_{c\vk}}$. 
\\
\indent
The BSE Hamiltonian in the electron-hole basis is
\begin{multline}
    \mathcal{H}_{I\!J}(\vq) = \Delta \epsilon_I(\vq)\, \delta_{I\!J} \\
       +\sqrt{\Delta f_I}(\vq)\left[\overline{V}_{I\!J}(\vq) - W_{I\!J}(\vq)\right]
        \sqrt{\Delta f_J(\vq)}.
    \label{eq:eh-hamiltonian}
\end{multline}
The interaction term in the second line includes the unscreened electron-hole exchange term, 
\begin{multline}    
    \overline{V}_{I\!J}(\vq) = \frac{1}{\Omega}\sum_{\mathbf{G}\neq 0} v(\vq+\vG)
    \rho_{cv\vk}(\vq+\vG)
    \rho_{v'c'\vk'}(\vq+\vG)
\label{eq:eh-exchange}
\end{multline}
and the screened electron-hole direct Coulomb term,   
\begin{multline}    
    W_{I\!J}(\vq) = \frac{1}{\Omega}\sum_{\mathbf{G},\mathbf{G}'}
    v(\vq_W + \mathbf{G})\epsilon^{-1}_{\mathbf{G},\mathbf{G}'}(\vq_W)\delta_{\mathbf{q}_W,\vk-\vk'} \\
    \rho_{cc'\vk}(\vq_W+\vG)
    \rho_{v'v\vk-\vq}(\vq_W+\vG),
\label{eq:eh-attraction}
\end{multline}
where $\Omega$ is the unit cell volume and $\vq_W$ is a shorthand for $\vk - \vk'$. The dipole matrix elements are defined as
\begin{equation}
\rho_{cv\vk}(\vq+\vG) = \delta_{\sigma_c,\sigma_v} \braket{c\vk|e^{i(\vG+\vq)\cdot\vecr}|v\vk-\vq}, 
\label{eq:rho-def}
\end{equation}
where spin conservation is a result of the bare Coulomb interaction being spin independent. 

The static dielectric function $\epsilon^{-1}_{\mathbf{G},\mathbf{G}'}(\mathbf{q}_W)$ in the direct term accounting for electron-hole attraction reads
\begin{equation}
    \epsilon^{-1}_{\mathbf{G},\mathbf{G}'}(\mathbf{q}) = \delta_{\vG,\vG'} + v_{\vG}(\vq)\chi_{\vG,\vG'}(\vq,\omega),
    \label{eq:dielectric-function}
\end{equation}
where the linear response function $\chi_{\vG,\vG'}(\vq,\omega)$ is obtained from the Dyson equation
\begin{multline}    
        \chi_{\vG,\vG'}(\vq,\omega) = \chi^0_{\vG,\vG'}(\vq,\omega)\\
        + \sum_{\vG_1,\vG_2}\chi^0_{\vG,\vG_1}(\vq,\omega)v_{\vG_1}(\vq)\delta_{\vG_1,\vG_2}\chi_{\vG_2,\vG'}(\vq,\omega).
\label{eq:x_dyson}
\end{multline}\\
\indent
For magnon calculations, the spin structure of the BSE Hamiltonian $\mathcal{H}_{I\!J}(\vq)$ is important. 
When SOC is neglected, $S_z$ becomes a good quantum number, and
we can label electron-hole transitions with pairs of spin indexes, $\{I\} \to \{\tilde{I}s\}$, 
where $s=\{\sigma_v,\sigma_c\}$ can take four different values:
$\{\uparrow\uparrow\}$, $\{\uparrow\downarrow\}$, $\{\downarrow\uparrow\}$, and $\{\downarrow\downarrow\}$, here called, respectively, 0, +, $-$, and 1. 
The BSE Hamiltonian matrix has the following spin structure~\footnote{Due to the conventions used in this paper, we label $H_{++}$ and $H_{--}$ the components that are typically referred to as $H_{+-}$ and $H_{-+}$ in the literature. This difference is a result of writing the BSE Hamiltonian in band-index space as $H_{cv,cv}$ instead of the usual $H_{vc,cv}$.}:
\begin{equation}
 \mathcal{H}_{I\!J}=
\left(
\begin{array}{cccc}
 H_{00} & H_{01} & 0 & 0  \\
 H_{10} & H_{11} & 0 & 0  \\
 0 & 0 & H_{++} & 0  \\
 0 & 0 & 0 & H_{--}
\end{array}
\right)
=
\begin{pmatrix}
\mathcal{H}^{E}_{I\!J} & 0 \\
0 & \mathcal{H}^{M}_{I\!J}
\end{pmatrix},
\end{equation}
where $\mathcal{H}^{E}_{I\!J}(\vq)$ is the standard excitonic BSE Hamiltonian for spin-conserving electron-hole transitions
and $\mathcal{H}^{M}_{I\!J}(\vq)$ is the magnon counterpart for spin-flip transitions. Both BSE Hamiltonians are $2 \times 2$ matrices in spin space. In non-magnetic systems, $\mathcal{H}^{E}_{I\!J}$ can be rearranged into singlet and triplet excitons with $S_z\!=\!0$ while $\mathcal{H}^{M}_{I\!J}(\vq)$ is block diagonal and describes two triplets with $S_z=\pm1$~\cite{Marsili2021}. Because the dipole defined in Eq.~\eqref{eq:rho-def} conserves spin, 
the electron-hole exchange term is zero in the spin-flip channel. When SOC is included, the two channels are mixed and one needs to compute the entire matrix $\mathcal{H}_{I\!J}(\vq)$.
\\
\indent

The BSE Hamiltonian can be used to obtain both optical and magnon properties, either by writing properties in terms of the Hamiltonian or by first diagonalizing it. 
To this end, we define the optical dipoles $d^i(\vq)$ (where $i$ labels Cartesian directions) and spin residuals $S^\pm(\vq)$ as vectors with components $I$ in transition space:
\begin{eqnarray}
d^i_I(\vq)=\frac{\rho_{cv\vk}(\vq)}{q} \sqrt{\Delta f_I(\vq)},
\label{eq:residuals1} \\ 
S^\pm_I(\vq)=\braket{c\vk|\sigma^\pm|v\vk-\vq} \sqrt{\Delta f_I(\vq)},
\label{eq:residuals2}
\end{eqnarray}
where $\sigma^{\pm}=\sigma_x\pm i\sigma_y$. The linear response of the system is contained in the
longitudinal polarizability~\cite{Onida2002}
\begin{equation}
    \alpha_{ij}(\w,\vq)=\langle d^i(\vq) | (\omega - \mathcal{H}(\vq))^{-1}| d^j(\vq) \rangle 
    \label{eq:alpha_def}
\end{equation}
and the spin susceptibility
\begin{equation}
    \chi^{+-}(\w,\vq) = \langle S^+ | (\omega - \mathcal{H}(\vq))^{-1}| S^- \rangle .
    \label{eq:chi_def}
\end{equation}

These expressions using the full BSE Hamiltonian $\mathcal{H}(\vq)$ apply to the general case where SOC is included. 
When SOC is neglected, $\alpha_{ij}(\w,\vq)$ is computed using
the exciton Hamiltonian $\mathcal{H}^{E}(\vq)$ in Eq.~\eqref{eq:alpha_def} and $\chi^{+-}(\w,\vq)$
is obtained using the magnon Hamiltonian $\mathcal{H}^{M}(\vq)$ in Eq.~\eqref{eq:chi_def}.
In both cases, these response functions can be  
computed efficiently using recursive techniques~\cite{Gruening2011}.\\
\indent
The exciton or magnon energies $E_{\lambda}(\vq)$, 
and corresponding wave functions $X^{\lambda\vq}_I$, are obtained 
by solving the eigenvalue problem for the respective Hamiltonians:
\begin{equation}
\mathcal{H}^{E/M}_{I\!J}(\vq)\, X^{\lambda\vq}_J = E_{\lambda}(\vq) X^{\lambda\vq}_I.
\end{equation}
This can be done with recursive approaches (for example, using the \textsc{SLEPc} library~\cite{Hernandez2005} as in \textsc{Yambo}) or by exact diagonalization. The advantage of using \textsc{SLEPc} is that one can compute only a subset of the full spectrum, with significant savings in memory and computational resources. This is particularly suitable for magnons, where only a few eigenvalues are tipically needed.

Starting from the exciton and magnon eigenvectors, the definition of correlated versions of the quantities in Eqs.~\eqref{eq:residuals1}-\eqref{eq:residuals2}, are admitted respectively
as $d^i_\lambda(\vq)=\sum_{I} X^{\lambda\vq}_I d^i_I(\vq)$ and
$S^\pm_\lambda(\vq)=\sum_{I} X^{\lambda\vq}_I S^\pm_I(\vq)$.
The exciton and magnon energies can be analyzed by defining the respective density of states (DOS):
\begin{eqnarray}
    D(\omega) &=& \sum_{\lambda\vq}\delta(\omega - E_{\lambda}(\vq)).
\end{eqnarray}

The exciton and magnon wave functions, $X^{\lambda\vq}_{cv\vk}$, can be analyzed in several ways. First, one can represent their amplitudes in frequency space~\cite{Galvani2016,Torun2018}:
\begin{equation}
  A^{\lambda\vq}(\omega) = \sum_{cv\vk}|X^{\lambda\vq}_{cv\vk}|^2\delta(\omega - (\epsilon_{c\vk}-\epsilon_{v\vq}))).
  \label{eq:a_omega}
\end{equation}

\noindent Second, they can be visualized in real space using
\begin{equation}
  \Psi_{\lambda\vq}(\mathbf{x}_e,\mathbf{x}_h) = \sum_{cv\vk}  X^{\lambda\vq}_{cv\vk}\, \psi^*_{c\vk}(\mathbf{x}_e)\,\psi_{v\vk-\vq}(\mathbf{x}_h).
  \label{eq:A_real_space}
\end{equation}
Finally, the wave functions can be analyzed in reciprocal space, either by plotting them as a function of electron momentum $\vk$ in the Brillouin zone or by mapping them on the band structure. This momentum-space representation is typically studied for $\vq\!=\!0$, but here we study it for finite transferred momentum $\mathbf{q}$, for electrons and holes separately using
\begin{subequations}
\begin{align}
  F_{c\vk}^{(e),\lambda\vq} = \sum_{v}|X^{\lambda\vq}_{cv\vk}|^2, \label{eq:Fe} \\
  F_{v\vk-\vq}^{(h),\lambda\vq} = \sum_{c}|X^{\lambda\vq}_{cv\vk}|^2.
  \label{eq:Fh} 
\end{align}
\label{eq:Feh}
\end{subequations}
These quantities allow the analysis of magnon states in momentum space. \\

\begin{figure*}[t]
\includegraphics[width=1.0\textwidth]{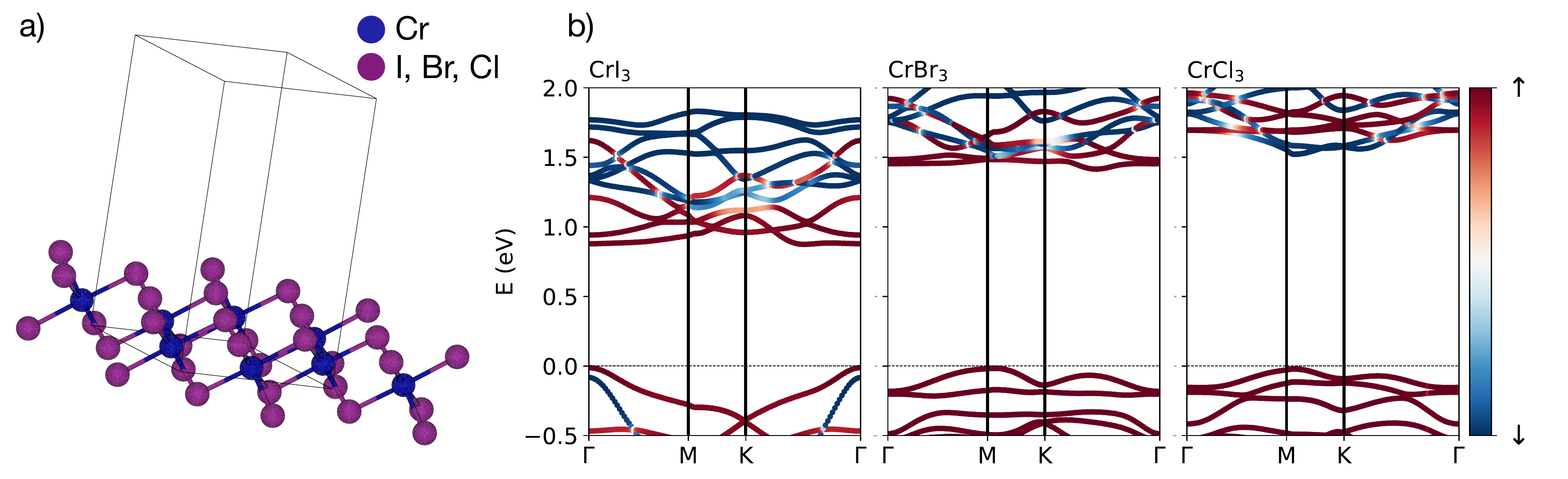}
  \caption{(a) Crystal structure of monolayer Cr trihalides. (b) Band structures of CrX$_3$ monolayers calculated using DFT with SOC. The color map shows the magnitude of the spin projection along the $z$-axis.} 
  \label{fig:crx3-structure}
\end{figure*}

\section{Methods}  \label{sec:methods}
We apply the BSE formalism to study magnons in CrX$_3$ monolayers.  The starting point for the BSE is the electronic structure, obtained here from DFT in a plane-wave basis as implemented in the \textsc{Quantum ESPRESSO} package~\cite{Giannozzi2009}. We compute the band structure of CrX$_3$ monolayers with and without SOC, using spinorial wave functions when SOC is included. The calculations use the local density approximation (LDA)~\cite{Perdew1981} and fully relativistic pseudopotentials generated using \textsc{ONCVPSP} \cite{Hamann2013} with semi-core valence electrons included for both Cr and X atoms. 
To simulate an isolated monolayer, we use unit cells with vacuum in the layer-normal direction, where the vacuum size is carefully converged to avoid spurious interactions with periodic replicas and we also apply a truncation of the Coulomb interaction in the $z$ direction. For all CrX$_3$ materials studied here, the lattice parameters are obtained from relaxation of the atomic coordinates and the lattice parameters with \textsc{Quantum ESPRESSO}. 
For magnon calculations, we correct the DFT band gaps using a scissor correction based on our previous GW calculations~\cite{Molina2020}. The magnon calculations are carried out using these corrected band structures.
The lattice parameters, scissor corrections, and other computational details are summarized in Table \ref{tab:DFT-conv-parameters}. 
\\
\indent
\begin{table}[b]
  \centering
  \resizebox{.49\textwidth}{!}
  {
  \begin{tabular}{| l | c | c | c |}
    \hline
    Halide & I & Br & Cl\\
    \hline
    a (\AA) SOC / no SOC & 6.77 / 6.70& 6.18 & 5.82\\
    \cline{2-4}
    c (\AA) & \multicolumn{3}{c|}{13.23}\\
    \hline
    Plane-wave energy cutoff (Ry) & 87 & 90 & 90\\
     \cline{2-4}
    k-mesh & \multicolumn{3}{c|}{12x12x1}\\
    \hline
    DFT gap no SOC (eV) & 1.29 & 1.50 & 1.55\\
     \cline{2-4}
    DFT gap SOC (eV) & 0.99 & 1.49 & 1.56\\
    \hline
    Scissor (eV) SOC / no SOC  & 1.87 / 1.47 & 2.96 / 2.94 & 3.92 / 3.93\\
    \hline
\end{tabular}
}
    \caption{Computational details for DFT calculations in CrX$_3$ monolayers, including lattice parameters, cutoff, energy gap, and scissor correction from GW calculations in Ref.~\cite{Molina2020}.}
    \label{tab:DFT-conv-parameters}
\end{table}
The BSE calculations, both with and without SOC, use a 12 Ry cutoff for the dielectric function and direct screened-Coulomb interaction, and an 11 Ry cutoff for the exchange interaction in the presence of SOC. (The electron-hole exchange interaction is zero without SOC). For calculations without SOC, we use the first 100 bands to compute the screening, and we construct the BSE Hamiltonian using 12 valence and 8 conduction bands; the number of bands is doubled for calculations with SOC. The DFT and BSE calculations employ the same $12\times 12\times 1$ $\mathbf{k}$-point grid
\footnote{This large grid is not needed to converge the ground state. However, to compute ground and excited states in a consistent way, we use the same grid for both calculations. This relatively dense grid provides reasonably well-converged magnon dispersions and absorption spectra in Cr$X_3$.}.
\\
\indent
The magnon energies are obtained from the poles of the spin susceptibility, $\chi^{\pm}$ in Eq. (\ref{eq:chi_def}), and scaled as described in Appendix. We enforce the Goldstone sum rule for acoustic magnons~\cite{Kotani2008,Muller2016,Olsen2021}, $\omega(\vq) \to 0$ for $\vq \to 0$, by adjusting the exchange splitting in the band structure, which equals the energy difference between the spin-up and spin-down bands involved in the magnon excitation~\cite{Muller2016}. 
Convergence of the BSE for magnons and excitons differs in important ways. While converging exciton energies typically requires relatively few bands and many $\mathbf{k}$-points, converging magnon energies requires many bands and relatively few $\mathbf{k}$-points. Overall, converging magnons is more challenging due to their smaller energy (meV for magnons versus order 100 meV for excitons) and the higher number of bands entering the BSE kernel. This point is further discussed in Appendix.

\section{Electronic and magnetic properties of chromium trihalides}  \label{sec:trihalides}
\vspace{-10pt}
In CrX$_3$ monolayers (X = I, Br or Cl), the Cr atoms are arranged in a honeycomb lattice and are at the center of the octahedra formed by halide atoms (Fig. \ref{fig:crx3-structure}(a)). These 2D materials are semiconductors with ferromagnetic order and their electronic and magnetic properties depend strongly on the halide atom \cite{Molina2020}: CrI$_3$ and CrBr$_3$ have point-group symmetry $D_{\rm{3d}}$ and out-of-plane magnetization, while CrCl$_3$ has magnetic point group $C_{\rm{2h}}$ and in-plane magnetization. For both this reasons and because the SOC in CrX$_3$ originates in the halide atoms, one expects that the magnon excitations will differ significantly for different halide atoms. 
\\
\indent
Figure \ref{fig:crx3-structure}(b) shows the band structures of monolayer CrX$_3$ materials. We find that the DFT band gap increases and the bands become less dispersive moving from lighter to heavier halide atoms along the CrCl$_3$ to CrI$_3$ sequence.  
The effect of SOC is strongest in CrI$_3$, where it leads to a clear spin mixing in the valence bands, while the effects of SOC are weaker in CrBr$_3$ and CrCl$_3$. 
\begin{figure*}[]
  \centering
  \includegraphics[width=1.05\textwidth]{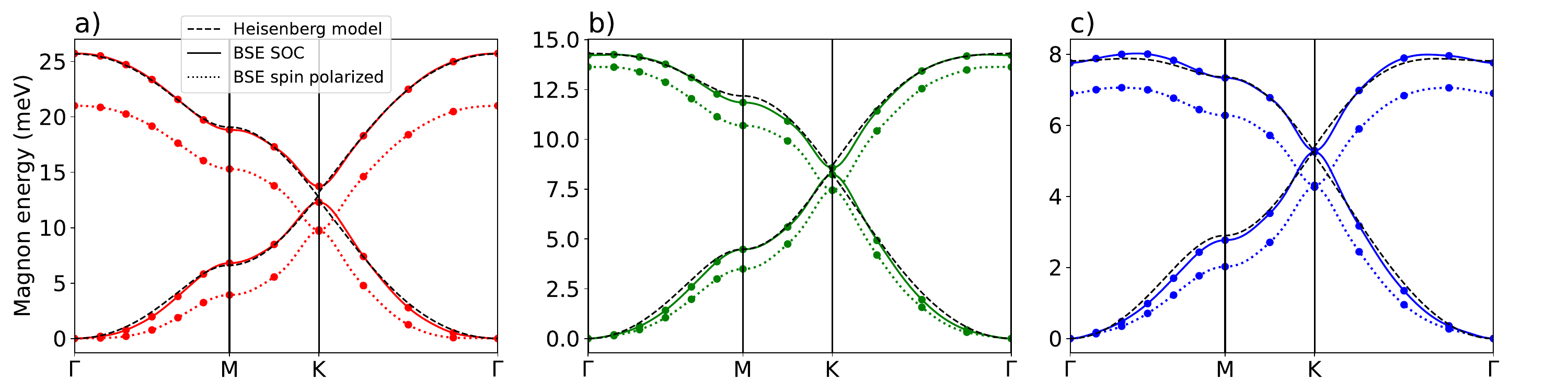}
  \caption{Magnon dispersions in monolayer trihalides: (a) CrI$_3$, (b) CrBr$_3$, and (c) CrCl$_3$, computed with SOC (solid lines) and without SOC (dotted lines). The points are BSE calculations, shown along with a smooth Fourier interpolation~\cite{Pickett1988} using continuous lines (SOC) or dotted lines (no SOC). The black dashed line is a fit to the Heisenberg model in Eq. \eqref{eq:heisenberg-hamiltonian}. The BSE results have been rescaled based on extrapolation to infinite cutoff and number of conduction bands, as discussed in the text. The same scaling factor was applied to calculations with and without SOC.} 
  \label{fig:crx3-model}
\end{figure*}
\\
\indent
The magnon dispersions in CrX$_3$ monolayers, computed with and without SOC, are shown in Fig. \ref{fig:crx3-model} along with the fits obtained from the Heisenberg model (see below).
The dispersions consist of an acoustic branch and an optical branch. The two branches have the largest magnon energy gap at $\Gamma$ and cross at the K point, where the magnon dispersion becomes linear, analogous to electronic Dirac cones in graphene \cite{Costa2020}.
The magnon energies decrease moving from heavier to lighter halide atoms (from CrI$_3$ to CrCl$_3$ in Fig.~\ref{fig:crx3-model}). This softening of magnon energies is accompanied by a decrease in the acoustic-to-optical magnon energy gap at $\Gamma$; values for this gap are given in Table \ref{tab:magnon-gaps}. This trend is a result of the different crystal and electronic structures for the three different compounds and not the SOC because the effect is also present in calculations without SOC. 
\\
\indent
When SOC is included, the magnon energies increase slightly and the magnon gap at K increases substantially (Table \ref{tab:magnon-gaps}). Because the halide atoms are responsible for SOC, the magnon gap at K becomes smaller for lighter halide atoms. For CrI$_3$, the calculated magnon gap at K is $\sim$1.5 meV. This value is greater than previous calculations using the BSE method~\cite{Olsen2021}, but is close to predictions from first-principles tight-binding calculations \cite{Costa2020}.
This point is discussed in more detail in Sec. \ref{subsec:heisenberg}.
\\
\vspace{20pt}
\indent

\begin{table}[b]
  \centering
  \begin{tabular}{| l | c c c || c c c |}
    \hline
    Halide & I & Br & Cl & I & Br & Cl\\
    \hline
    $\Delta^{\rm{noSOC}}_\Gamma$ (meV) & 25.8 & 16.7 & 8.5 & 21.0 & 13.6 & 6.93 \\
    $\Delta^{\rm{SOC}}_\Gamma$  (meV) & 31.5 & 17.4 & 9.5 & 25.7 & 14.2 & 7.7 \\
    $\Delta^{\rm{SOC}}_K$ (meV) &  
     1.8 &  0.4 & 0.0 & 1.5 & 0.3 & 0.0\\
    \hline
    Convergence & \multicolumn{3}{c||}{Raw} & \multicolumn{3}{c|}{Extrapolated} \\
    \hline
  \end{tabular}
  \caption{Energy difference, $\Delta_\vq$ in meV units, between the acoustic and optical magnon branches, given at $\vq=\Gamma$ and $\vq=K$ for the CrX$_3$ monolayers studied here. The values are obtained by diagonalizing the stable BSE (see Appendix) with and without SOC, and extrapolated to infinte parameters.}
  \label{tab:magnon-gaps}
\end{table}
We analyze our results by fitting the BSE magnon dispersions to a Heisenberg model:
\begin{equation}
    H = -\frac{1}{2}\sum_{\braket{ij}}J_{ij}\mathbf{S}_i\cdot\mathbf{S}_j,
\label{eq:heisenberg-hamiltonian}
\end{equation}
\noindent where $J_{ij}$ is the exchange interaction between spins $i$ and $j$. 
We employ the \textsc{SpinW} code to construct and diagonalize this Heisenberg Hamiltonian~\cite{Toth2015}. The \textit{ab initio} BSE results are well reproduced with isotropic exchange parameters up to the third nearest neighbor (Fig. \ref{fig:crx3-model}). Table \ref{tab:spinw-params} gives the exchange parameters for the three Cr trihalides studied here, where the second nearest-neighbor exchange parameter is zero for CrI$_3$ and very small for the other trihalides. Our exchange parameters are slightly different from previous reports, as we discuss in Sec.~\ref{subsec:heisenberg}.   
Overall, the agreement between magnons from the BSE and the Heisenberg model is very good, especially considering that in the Heisenberg model the magnons originate entirely from the
magnetic moment of Cr atoms, with no contribution from halide atoms. Note that the Heisenberg model cannot predict the magnon gap at $K$, which requires including SOC~\cite{Zhang2022}. 

\begin{table}[b]
  \centering
  \begin{tabular}{| l | c c c || c c c |}
    \hline
    Halide & I & Br & Cl & I & Br & Cl\\
 \hline
 
    $J_1$ (meV) & 1.950 & 1.125 & 0.628  & 1.590 & 0.918 & 0.512\\
    $J_2$ (meV) & 0.000 & 0.060 & 0.063& 0.000 & 0.049 & 0.051\\
    $J_3$ (meV) & -0.200 & -0.150 & -0.095& -0.163 & -0.122 & -0.078\\
    \hline
    Convergence & \multicolumn{3}{c||}{Raw} & \multicolumn{3}{c|}{Extrapolated} \\
    \hline
\end{tabular}
    \caption{Exchange parameters for the Heisenberg model in Eq.~\eqref{eq:heisenberg-hamiltonian} for the CrX$_3$ monolayers studied here.} 
    \label{tab:spinw-params}
\end{table}

\section{Analysis of magnons in C\MakeLowercase{r}I$_3$}  \label{sec:detailscri3}
\begin{figure}[t]
  \centering
  \includegraphics[width=0.45\textwidth]{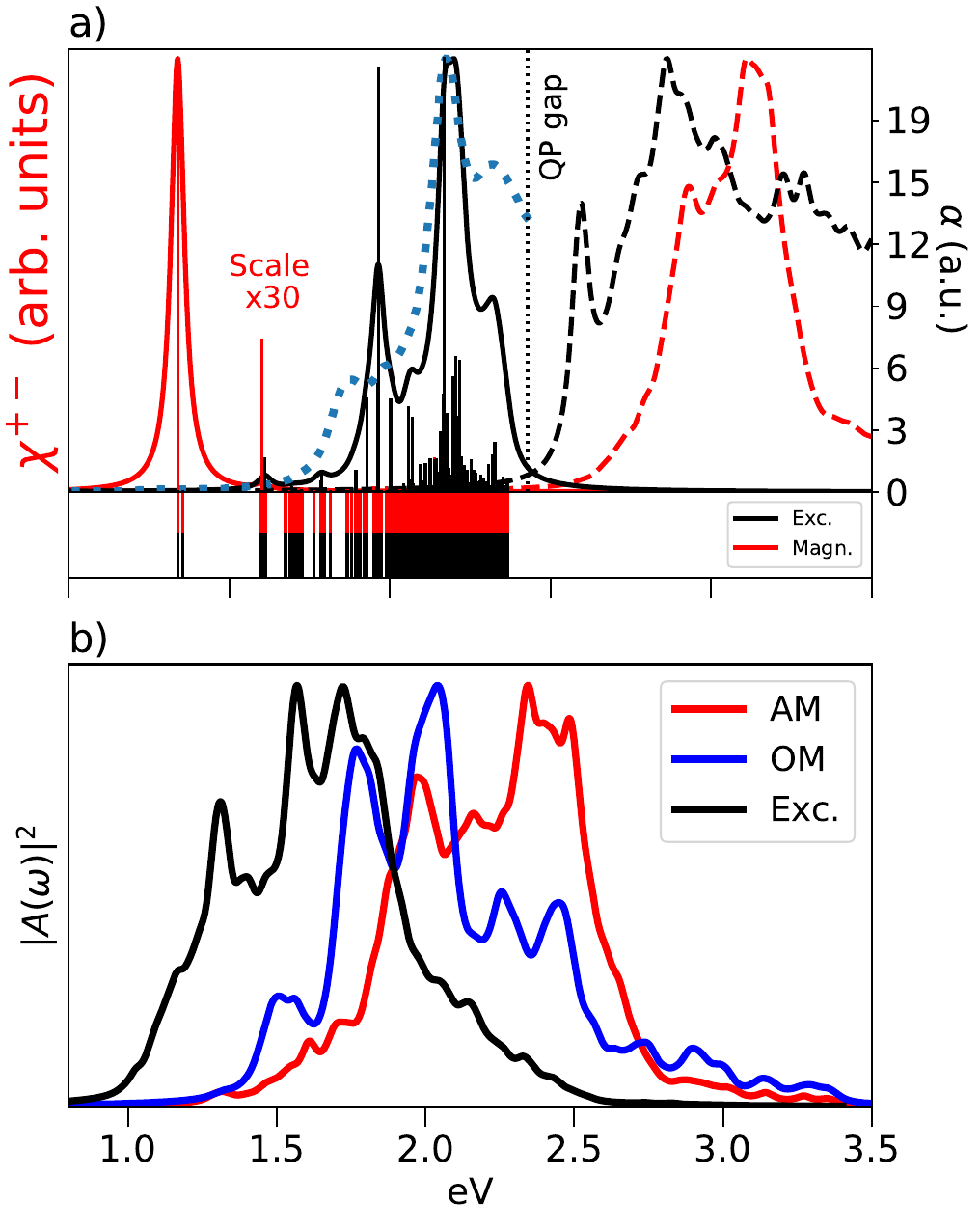}
  \caption{(a) Spin susceptibility (red) and polarization (black) at $\vq = 0$ in CrI$_3$, computed with the BSE (solid lines) or using the independent-particle approximation (dashed lines), in both cases with SOC. The peaks of the two response functions are normalized to same height. 
  The broadening values for the spin susceptibility are 1 meV (BSE) and 50 meV (independent-particle calculation). The polarization employs a 50 meV broadening.
  The relative weights of the eigenvalues are shown as vertical black (excitons) and red lines (magnons). Magnon weights are scaled with a $x30$ factor with respect to the first magnon peak. Also shown for comparison is the optical absorption from Ref.~\cite{Olsen2021} (blue dotted line). The BSE eigenvalues in the spin-conserving and spin-flip channels, labeled respectively as magnons and excitons, are shown as vertical lines in the lower inset.
  (b) Square amplitude $|A(\omega)|^2$, with $|A(\omega)|^2$ defined in Eq.~\eqref{eq:a_omega}, for acoustic magnons (AM), optical magnons (OM), and lowest-energy bright exciton (Exc.).
  \label{fig:cri3-peaks-soc}
  }
\end{figure}

\vspace{-10pt}
In this work, we compute magnons using the \textit{ab initio} BSE, which is typically employed to study excitons. Yet, magnons and excitons differ in important ways. 
Magnons are collective spin excitations with canted spin orientations at the magnetic-atom sites, while excitons are bound electron-hole pairs, typically classified into Frenkel or Wannier excitons depending on their spatial localization. The BSE provides a common approach to study excitons and magnons, allowing us to compare and contrast these two excitations.
\\
\indent
For this discussion, we focus on the representative case of CrI$_3$ with SOC. 
We analyze the polarizability $\alpha(\omega)$ and the spin susceptibility $\chi^{+-}(\omega)$, defined respectively in Eqs. (\ref{eq:alpha_def}) and (\ref{eq:chi_def}). 
Figure~\ref{fig:cri3-peaks-soc}(a) shows these two response functions, in both cases comparing calculations with the BSE method, which includes electron-hole interactions, and using the independent particles (IP) approximation, where electron-hole interactions are neglected~\cite{Onida2002}. As the calculations include SOC, we obtain a common set of BSE eigenvalues for both response functions, but some eigenvalues contribute only to $\chi_{+-}(\omega)$ and others only to $\alpha(\omega)$. 
\\
\indent
Based on the behavior of the spin-susceptibility, we refer to the two lowest-energy eigenvalues as ``magnons''. These two BSE solutions, which correspond to acoustic and optical magnon excitations, are well-separated in energy from the rest of the spectrum. The acoustic magnon dominates the intensity of $\chi^{+-}(\vq,\omega)$, with $\vq$ in the first Brillouin zone, while the optical magnon dominates the intensity of $\chi^{+-}(\vq+\vG,\omega)$, being $\vG$ any reciprocal lattice vector that connects two Cr atoms \cite{Olsen2021}. These two excitations were also found in previous BSE calculations, although they were initially referred to as dark spin-flip excitons~\cite{Wu2019} and only later were addressed as magnon excitations in Ref. \cite{Olsen2021}. Since magnons are Goldstone modes emerging from breaking rotational symmetry, their energy should vanish in the $\vq \to 0$ limit. 
However, in our calculation, the magnon energy in CrI$_3$ is nonzero due to a well-known violation of the Goldstone sum rule in GW-BSE magnon calculations.  Although rigorous solutions to this technical point have been proposed~\cite{Muller2016}, here the magnon dispersions in Fig.~\ref{fig:crx3-model} are obtained simply by shifting the magnon energies down by the calculated BSE magnon energy at $\vq = 0$, which is 1.25 eV in our CrI$_3$ results. 
\\
\indent
We refer to the rest of the eigenvalues below the quasiparticle gap as ``excitons''. Some of these excitations are optically active (bright) and contribute to the computed polarization response shown in Fig.~\ref{fig:cri3-peaks-soc}(a), which agrees with previous results~\cite{Wu2019,Olsen2021}. The other excitations are not optically active (dark), either because they consist of spin-flip excitations or because the dipole matrix elements vanish due to symmetry. 

To analyze the nature of magnons and excitons, in Fig.~\ref{fig:cri3-peaks-soc}(b) we also plot the square of the amplitudes $A^{\lambda \vq}(\omega)$ defined in Eq.~\eqref{eq:a_omega}, which quantify the contribution of electronic transitions at energy $\omega$ to the magnon or exciton state $\lambda$, calculated here for vertical electronic transitions ($\vq=0$). 
We find that both optical and acoustic magnons result from electronic transitions spanning a wide energy range from about 1 $-$ 3.5 eV. 
In contrast, the lowest-energy bright exciton state, which is 0.7 eV higher than the magnon eigenvalues, has contributions from transition energies up to only 2.5 eV, and thus 1 eV smaller than for magnons. 
\\
\indent
For excitons in 2D semiconductors, this is still a relatively large energy range of contributing electronic transitions. Since Wannier excitons in 2D materials are typically contributed by a smaller energy range of electronic transitions~\cite{Molina2013}, our results suggest the presence of a localized Frenkel-like exciton in monolayer CrI$_3$. 
The magnon is associated with an even wider transition energy range, leading to a magnon wave function localized at the magnetic (Cr) atom sites and to a greater BSE binding energy for magnons compared to the lowest-energy exciton. 
\begin{figure}[t]
  \centering
\includegraphics[width=0.49\textwidth]{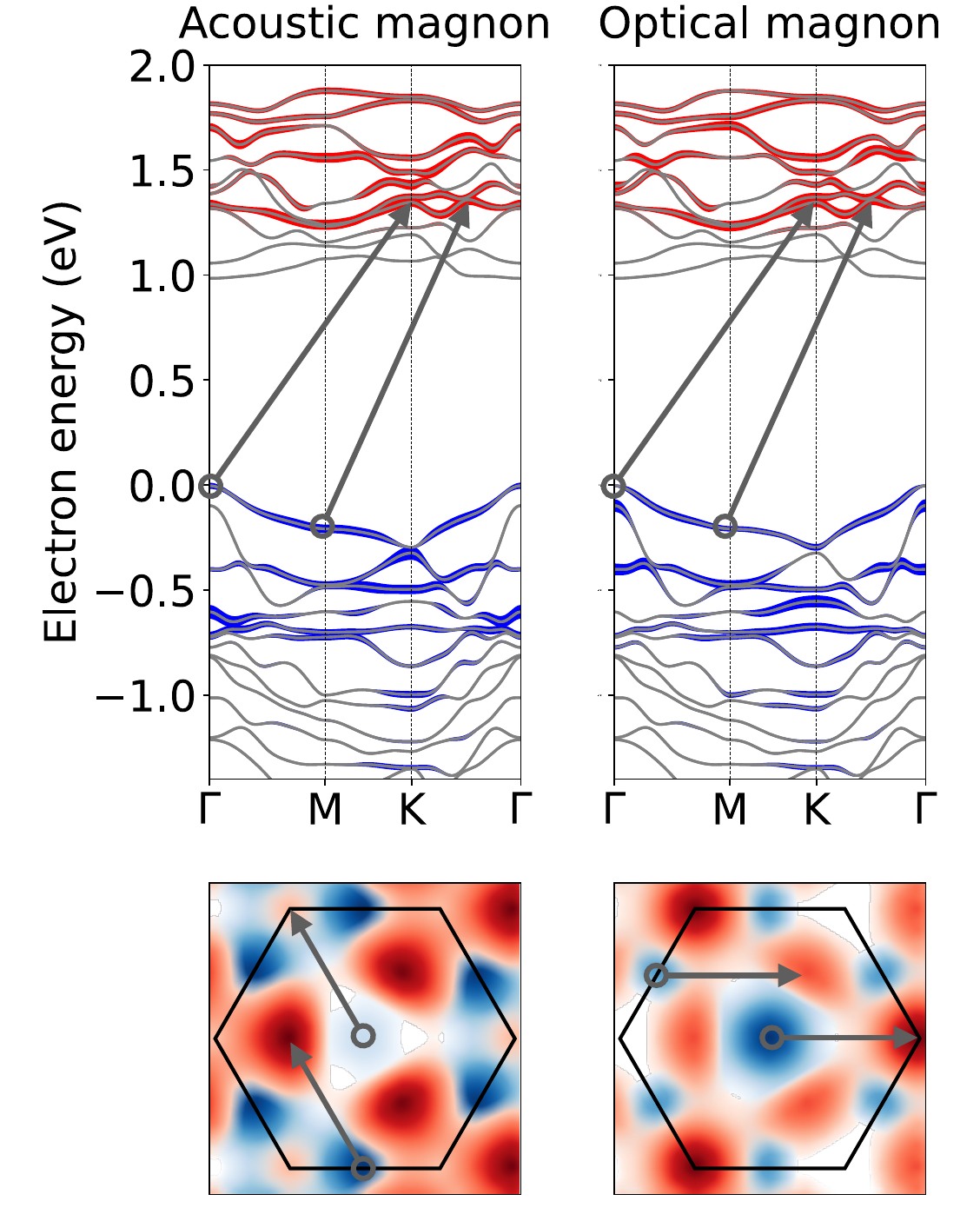}
\caption{(a) Electronic transitions contributing to acoustic and optical magnons at $\mathbf{q} = K$, quantified by plotting the weights $F_{c\vk}$ for electrons (red) and $F_{v\vk-\vq}$ for holes (blue) defined in Eqs.~\eqref{eq:Fe} and \eqref{eq:Fh}. (b) Two-dimensional
map of magnon weigths in the hexagonal Brillouin zone. We show only weights larger than 95\%. In both panels,the gray arrows show transitions from the valence to the conduction band.}

\label{fig:cri3-magnons-soc-bz-bands}
\end{figure}

\subsection{Magnon wave functions}  \label{subsec:wavefunc}
\vspace{-10pt}
We analyze the magnon wave functions by studying its localization in both real and reciprocal space. 
We first focus on the localization of the electron relative to the hole, which can be studied by plotting the electron wave function for a fixed hole position, or alternatively by analyzing the wave function in momentum space. This analysis is common for BSE calculations of excitons but not for magnons. 
\\
\indent
As discussed above, the BSE magnon wave function in CrI$_3$ has contributions from electron-hole transitions with a wide energy range. Therefore, we expect that these transitions that make up the magnon wave function are delocalized over the entire Brillouin zone in reciprocal space. We employ the momentum-space representation of the BSE wave function introduced in Eqs. \eqref{eq:Fe}-\eqref{eq:Fh}, which is useful for mapping the contributions to the magnon wave function on the electronic band structure. Figure~\ref{fig:cri3-magnons-soc-bz-bands}(a) shows this analysis for acoustic and optical magnons in CrI$_3$ with momentum $\vq=K$. For both magnon modes, we find that the BSE wave function extends almost uniformly over the entire Brillouin zone and across several valence and conduction bands, consistent with the collective nature of magnon excitations. 
\\
\indent
Despite this similarity, even though these two modes are nearly degenerate, there are important differences between the acoustic and optical magnons at $\mathbf{q}=K$.  
Mapping the reciprocal-space wave functions in the 2D Brillouin zone of CrI$_3$ (Fig. \ref{fig:cri3-magnons-soc-bz-bands}(b)) reveals that the main contribution to the optical mode are electronic transitions from $\Gamma$ to $K$, and for the acoustic mode from $M$ to momenta between $\Gamma$ and $K$. In addition, the acoustic magnons have stronger contributions from the highest valence bands and the optical magnons from lower-energy valence bands. 
\\

\begin{figure*}[!t]
\centering
\includegraphics[width=0.8\textwidth]{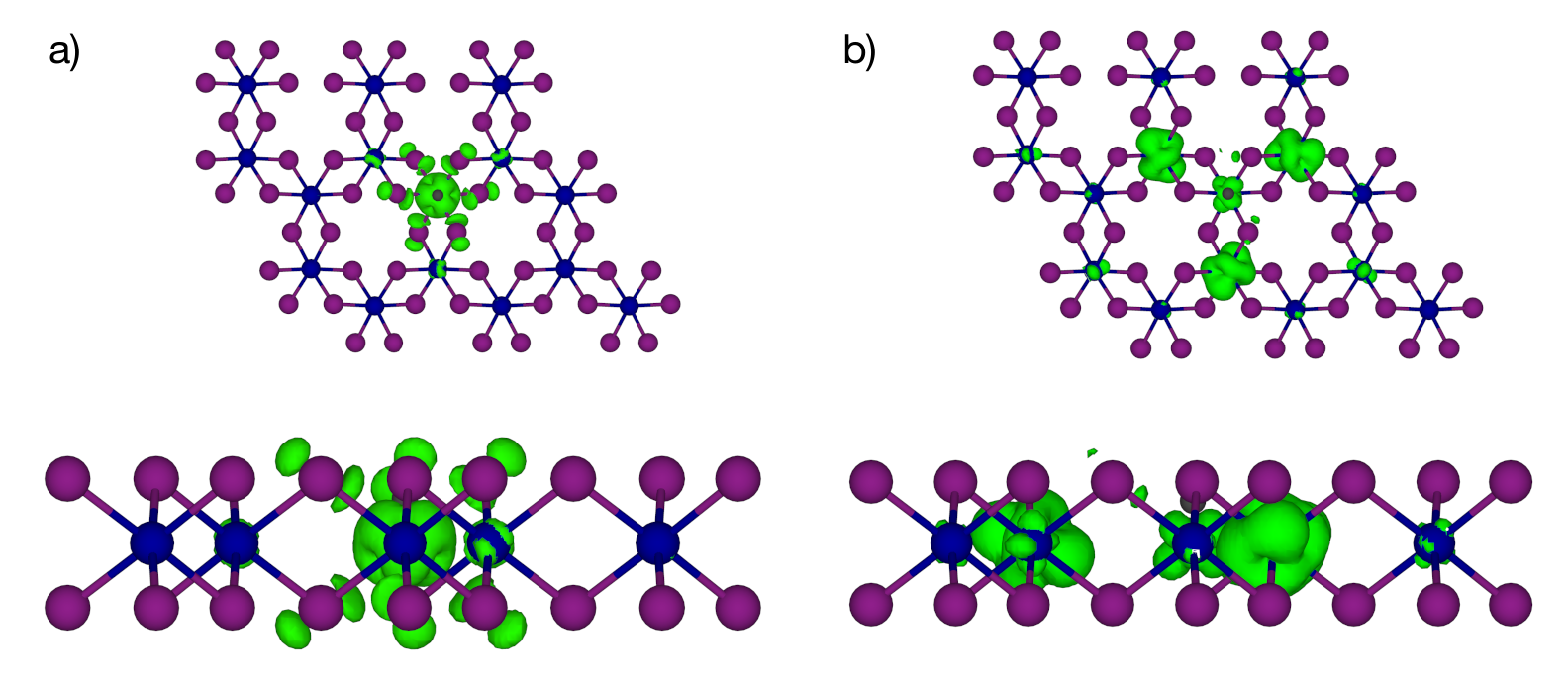}
\caption{Magnon wave functions in real space obtained from the BSE with SOC, studied by plotting the electron probability density $n_{\lambda \vq}(\mathbf{x}_e)$ with the hole placed slightly above the Cr atom in the center of the structure. Results are given for (a) acoustic magnon $\mathbf{q} = K$ and (b) optical magnon at $\mathbf{q} = K$. The plots show iso-surfaces at 0.38 \% of the maximum amplitude.}
\label{fig:cri3-real_space}
\end{figure*}

\indent
This delocalization in reciprocal space leads to localized magnon wave functions in real space. This is shown in Fig. \ref{fig:cri3-real_space}, where we plot the magnon wave functions in real space using Eq.~\eqref{eq:A_real_space}. Following the standard analysis of BSE excitons~\cite{Sangalli2019}, we fix the hole position at $\mathbf{x}^{*}_h$, which in this case is near one of the Cr atoms, and compute the electron density around the hole, $n_{\lambda \vq}(\mathbf{x}_e) = |\Psi(\mathbf{x}_e, \mathbf{x}^{*}_h) |^2$, for select magnons with mode index $\lambda$ and wave-vector $\vq$. 
(Different from excitons, in the magnon case the electron and hole have opposite spins, and thus $n_{\lambda \vq}(\mathbf{x}_e)$ quantifies the probability of finding a minority-spin electron in the conduction band given the presence of a majority-spin hole in the valence band.) 
\\
\indent
The wave function for acoustic magnons at $\mathbf{q} = K$ is localized on the Cr atom, with a significant density near the halide atoms (Fig. \ref{fig:cri3-real_space} (a)). This trend is a departure from a simple Heisenberg model, which assumes that spins are localized at the magnetic atoms. Our results are somewhat different from previous work~\cite{Wu2019}, where the magnon wave function is fully localized on the Cr atoms. We attribute this difference to our choice of not placing the hole exactly at the Cr atom, where the wave function has a node, but rather very close to it. 
The real-space behavior of optical magnons is qualitatively different from acoustic magnons. Figure~\ref{fig:cri3-real_space}(b) illustrates this result for the optical magnon at $\mathbf{q} = K$. Relative to the Cr atom where the hole is located, the electron density $n_{\lambda \vq}(\mathbf{x}_e)$ is small near the Cr-atom hole site and greater at the three nearest-neighbor Cr atoms, with negligible density on the halide atoms. This clearly shows that the nearly degenerate acoustic and optical magnons at $\mathbf{q} = K$ are qualitatively different magnetic excitations.
\\
\indent
Our results show that electrons and holes are localized on the same site or at nearest-neighboring sites, suggesting a rapidly decaying exchange interaction between Cr atoms with main contributions from first nearest neighbors. This picture is consistent with the exchange coupling parameters extracted from our BSE magnon dispersion (Table~\ref{tab:spinw-params}), where the nearest-neighbor exchange coupling $J_1$ is an order of magnitude greater than the second and third neighbor values $J_2$ and $J_3$. We have verified that the wave functions for acoustic magnons at $\Gamma$ and at other momenta show a similar real-space behavior. 

\subsection{Role of lattice parameter and extrapolation}
\label{subsec:lattice} 
\vspace{-10pt}

\begin{figure}[b]
  \includegraphics[width=0.49\textwidth]{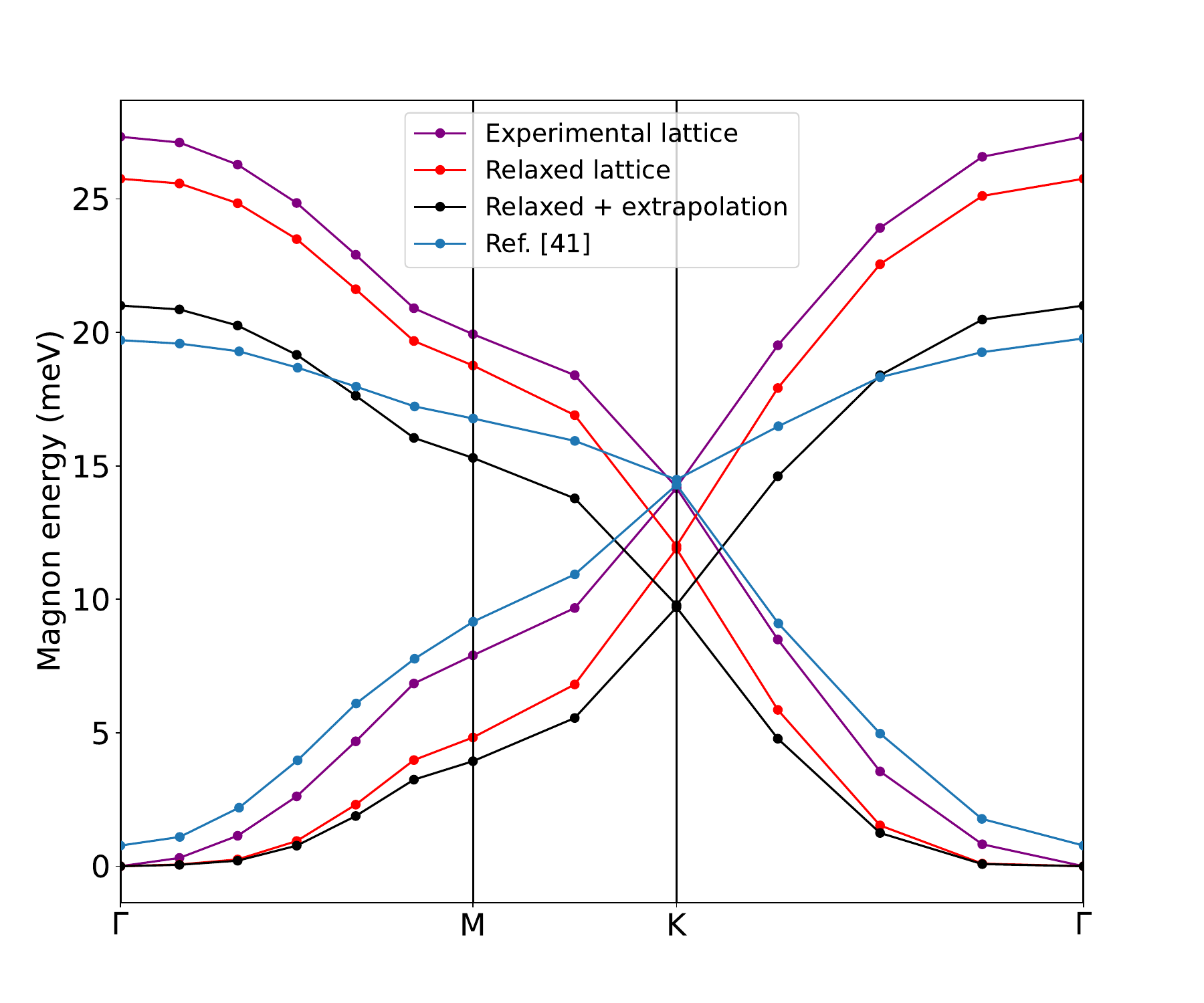}
  \caption{Magnon dispersions in CrI$_3$ computed using, respectively, the BSE with relaxed lattice parameters, shown separately with or without extrapolation, and the BSE with experimental lattice constants. The magnon dispersion from Ref.~\cite{Olsen2021} is shown in blue for comparison. 
  }
  \label{fig:cri3_bse_magnons}
\end{figure}

The magnon dispersions computed with the BSE are somewhat sensitive to the choice of lattice parameters and extrapolation procedure~\cite{Esteras2023}. The BSE calculations presented above are carried out with lattice parameters relaxed with DFT. To study the effect of lattice parameters on magnon dispersions, we also perform BSE calculations on CrI$_3$ using experimental parameters~\cite{Villars2023}. The results for relaxed and experimental lattice parameters are compared in Fig.~\ref{fig:cri3_bse_magnons}.  
Using experimental lattice parameters does not change the main features of the magnon dispersions and only leads to an overall stretching of the dispersion curves relative to results for relaxed lattice parameters. For example, the acoustic to optical gap at $\Gamma$ changes from 26 to 28 meV, which is a stretching of $\sim$8\%.  
\\
\indent
The impact of the extrapolation procedure (see Appendix) is more pronounced, as we show in Fig.~\ref{fig:cri3_bse_magnons} by comparing results with and without extrapolation using relaxed lattice parameters. When extrapolation is used, the acoustic-to-optical gap at $\Gamma$ changes from 26 to 21 meV. 
We also compare our results with the magnon dispersions from Ref.~\cite{Olsen2021}, which also employs the BSE with an extrapolation procedure. The two sets of magnon dispersions are in reasonable agreement with each other $-$ in particular, the acoustic-to-optical gap at $\Gamma$ is very similar in the two calculations, with values of 19 meV in Ref.~\cite{Olsen2021} and 21 meV in our calculations with extrapolation. The energy of the crossing between the acoustic and optical magnons at $K$ is somewhat different in the two calculations, with values of 14 meV in Ref.~\cite{Olsen2021} versus 10 meV in this work. Despite the inherent variability of BSE results with respect to lattice parameters, cutoffs, and extrapolation, our results show that BSE calculations are overall consistent across different computational settings and implementations.

\subsection{Comparison with time-dependent DFT}
\label{subsec:tddft}
\begin{figure}[!ht] \includegraphics[width=0.49\textwidth]{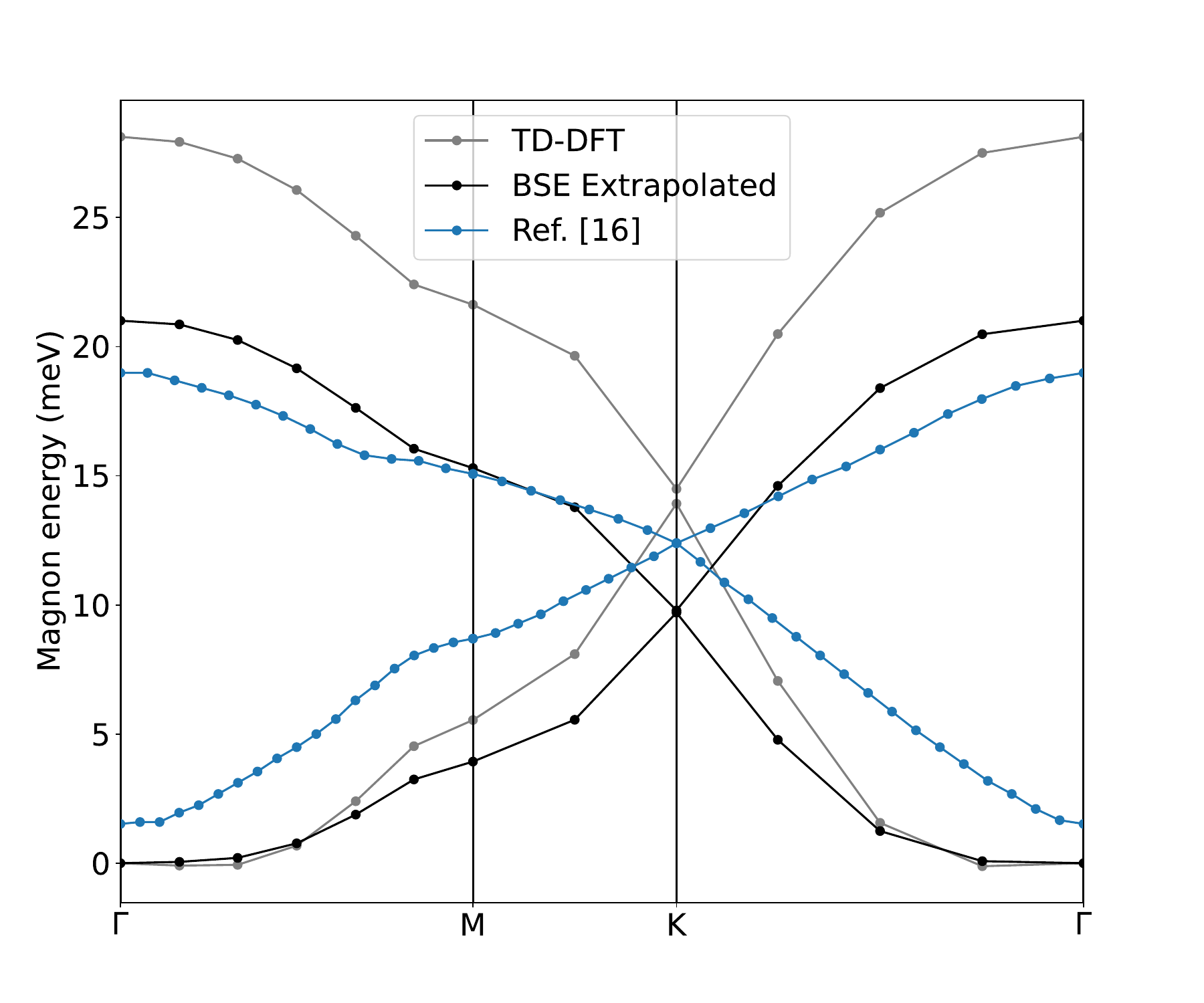}
  \caption{Comparison between magnon dispersions in CrI$_3$ computed with the TD-DFT and BSE methods. The two calculations using similar settings, including relaxed lattice constants, the same number of bands, and converged kernel cutoffs in both cases. The two magnon dispersions have been manually shifted to zero energy at $\Gamma$ to satisfy the Goldstone sum rule. The experimental magnon dispersion for CrI$_3$ from Ref. \cite{Chen2018} is included in blue.}
  \label{fig:cri3_tddft-bse-raw}
\end{figure}

There is a range of empirical and first-principles methods available for computing magnon dispersions. For completeness, we compare our BSE results with time-dependent DFT (TD-DFT) magnon calculations, also implemented in the \textsc{Yambo} code~\cite{Sangalli2019}. To make a fairer comparison, we use the same crystal structure and number of bands in both calculations, in each case using a converged kernel cutoff value (the kernel cutoff at convergence is \mbox{90 Ry} in TD-DFT and 11 Ry in the BSE).
\\
\indent
Figure \ref{fig:cri3_tddft-bse-raw} represents a comparison between results from BSE and TD-DFT in CrI$_3$. The magnon dispersions from the two methods are in very good agreement with each other, with only a small difference (2.5 meV) in the acoustic-to-optical magnon gap at $\Gamma$ and a relative shift of the magnon spectra by about 2 meV for the mode-crossing at $K$. In addition, note that the Goldstone sum rule violation occurs in both our BSE and TD-DFT calculations. Including a larger number of bands may alleviate this issue in TD-DFT, although the wide energy range of electronic transitions contributing to magnons in CrI$_3$ may slow the convergence with respect to number of bands.
\\
\indent
These results show that the finite-momentum BSE is a reliable first-principles approach for quantitative studies of magnon excitations in real materials. It is a valid alternative to TD-DFT for magnon calculations with reliable kernel interactions.\\
\subsection{Heisenberg exchange parameters}  
\label{subsec:heisenberg}
\begin{figure}[!t]
  \centering
  \includegraphics[width=0.49\textwidth]{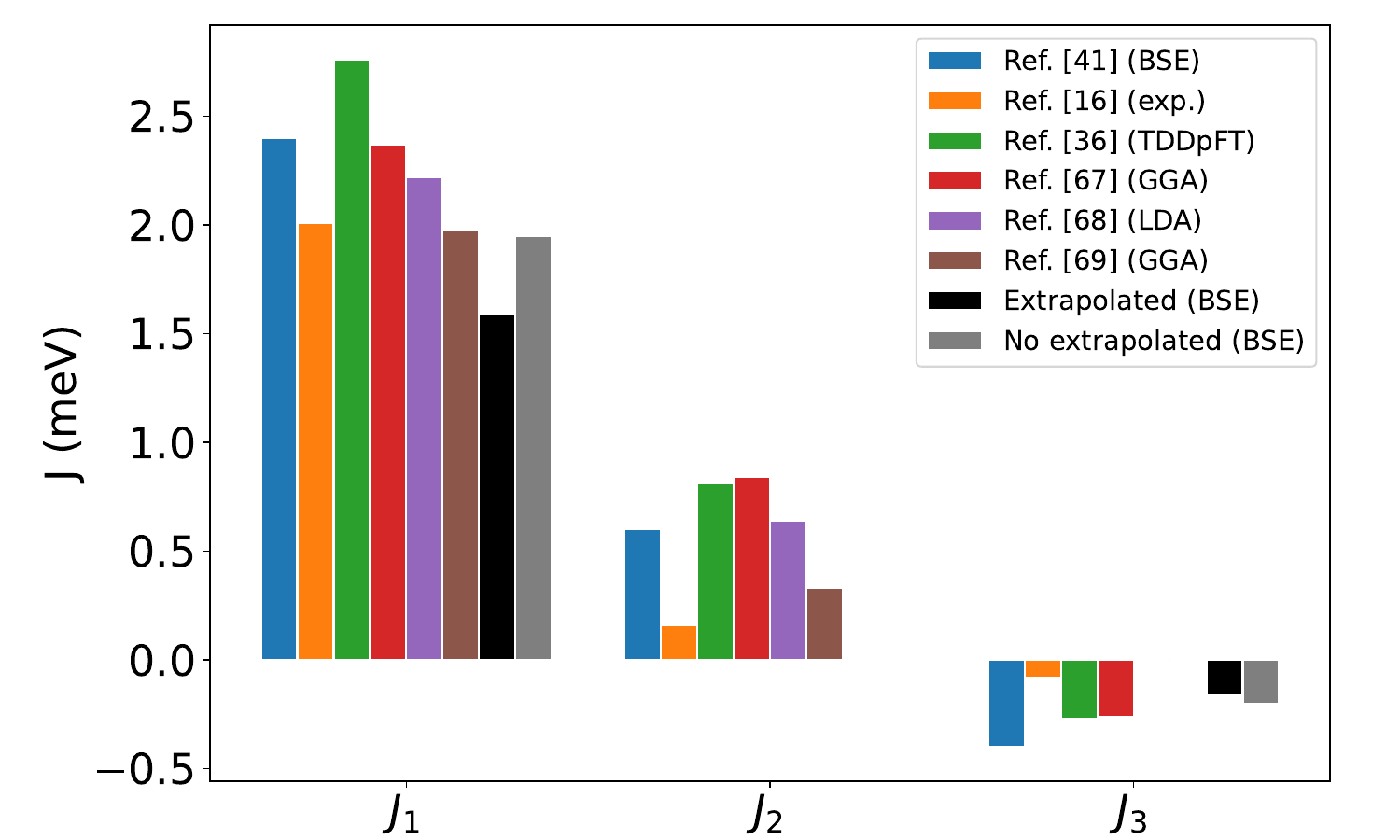}
  \caption{First, second, and third nearest-neighbor exchange coupling parameters, $J_1$ to $J_3$, for the Heisenberg model in Eq.~\eqref{eq:heisenberg-hamiltonian}. Values from the literature are compared with those obtained here from the BSE with or without extrapolation.}
  \label{fig:exchange-scheme}
\end{figure}

\vspace{-10pt}
The Heisenberg model for Cr trihalides typically includes exchange coupling parameters $J_1$, $J_2$ and $J_3$, respectively for exchange coupling between first, second, and third nearest-neighbor Cr atoms. 
The exchange parameters available in the literature for magnons in CrI$_3$ span a wide range of values~\cite{Chen2018,Besbes2019,Xu2020,Olsen2021,Delugas2023,Soenen2023}. These parameters are obtained with several methods, including DFT, TD-DFT, and BSE calculations in monolayer CrI$_3$, or by fitting experimental data from bulk samples. 
We summarize the exchange parameters from different sources in Fig.~\ref{fig:exchange-scheme} and compare them with our \mbox{BSE calculations.}
\\
\indent
All results agree on the ferromagnetic nature of second-neighbor exchange and antiferromagnetic third-neighbor exchange ($J_2>0$ and $J_3<0$, respectively). However, the exchange values from different sources can be significantly different. For example, our calculated value of $J_1$ is close to the DFT-GGA calculations from~\cite{Soenen2023}, but collectively our three values $J_i$ are closest to the exchange parameters from experimental fits in bulk samples~\cite{Chen2018}. 
The DFT-LDA results from Ref.~\cite{Olsen2021} agree on $J_1$ with the DFT-GGA calculations from Ref.~\cite{Besbes2019}, but differ significantly for the other two exchange couplings. In addition, the $J_2$ and $J_3$ values obtained with TD-DPT in Ref.~\cite{Delugas2023} are nearly identical to the corresponding values in Ref.~\cite{Besbes2019}. Finally, the relative values of $\{J_1,J_2,J_3\}$ in Refs.~\cite{Olsen2021} and \cite{Delugas2023} are in excellent agreement with each other, although with slightly different absolute values. 
\\
\indent
To illustrate the effects of these discrepancies, note that the shape of the magnon dispersion from the Heisenberg model is determined by the relative values of the couplings $J_{i}$. The energy difference between the acoustic and optical modes at $\Gamma$ is proportional to the sum of inter-site couplings $(J_1 + J_3)$, while the relative curvature of the acoustic and optical branches depends on the ratio $J_2/(J_1+J_3)$. In the limit of $J_2 \to 0$, the two branches show equal and opposite dispersion about the mid-point of the gap at $\Gamma$, and the curvature of both branches depends on the ratio $J_1/J_3$. 
In this limit, when $J_1$ and $J_3$ have the same sign the curvature of the optical branch will always be negative, and when they have opposite signs the curvature will depend on the absolute value of the ratio. Yet, imposing $J_2 \!=\! 0$ can lead to negative acoustic branches for large $J_1 / J_3$ ratio, in which case it is preferable to break the mirror symmetry by imposing $J_2 \neq 0$.
\\
\indent
Given this extensive interplay of exchange parameters, one expects a wide variability in computed magnon dispersions in CrI$_3$. In particular, different papers disagree on the curvature of the optical branch along $\Gamma$ to $M$, with Refs.~\cite{Olsen2021,Delugas2023,Besbes2019,Xu2020} showing an optical branch with positive curvature, while the curvature is negative in our work as well as previous BSE calculations~\cite{Olsen2021}, experiments~\cite{Chen2018}, tight-binding calculations~\cite{Costa2020}, and results in Ref.~\cite{Soenen2023}. 
\\
\indent
We extend this analysis to CrBr$_3$ and CrCl$_3$, and find a good agreement between our computed values $J_i$ and experimental reports for CrBr$_3$~\cite{Cai2021}, and less satisfactory agreement with experimental data for CrCl$_3$~\cite{Chen2021} although our computed value of $J_1$ are close to recent post-Hartree-Fock calculations \cite{Yadav2024}. We also find slight differences in the curvature and acoustic-to-optical gap at $\Gamma$ for CrCl$_3$ compared to recent work~\cite{Brehm2024}. We believe that the rapidly developing field of magnons in 2D materials will benefit from such detailed comparisons between different results and methods. 

%
\section{Conclusions}  \label{sec:conclusions}
\vspace{-10pt}
This work shows \textit{ab initio} calculations of magnon dispersions and wave functions in monolayer Cr trihalides using the framework of the BSE, which accounts for electron-hole spin correlations and SOC effects. 
Our results clarify the dependence of magnon dispersions on the halide atoms and the strength of SOC. Lighter halide atoms are associated with lower magnon energies, and SOC is responsible for opening a gap between the acoustic and optical magnon branches at $K$. The size of this gap increases with increasing SOC strength, and the gap nearly vanishes for the material with the lightest halide atom, CrCl$_3$. These results are robust against different choices of lattice 
parameters. The convergence is numerically very demanding, and it could 
be achieved via extrapolation techniques. While the final result is not 
qualitatively different from converged TD-DFT calculations, 
quantitatively we see a softening of the dispersion when comparing BSE 
to TDDFT, which is in better agreement with available experimental data.
Our results show that a Heisenberg model with isotropic exchange interactions can satisfactorily reproduce the magnon dispersions, although more general models are needed to capture the magnon gap at $K$. The exchange parameters extracted from the BSE agree with previous reports for CrI$_3$ and CrBr$_3$ but are slightly different from measured values in CrCl$_3$. Our analysis highlights the wide range of Heisenberg exchange parameters obtained with different approaches. Building on these and other technical advances, we developed first-principles magnon-phonon calculations in the companion paper \cite{companion}. Future work will apply the BSE method more broadly to study magnons in 2D and bulk materials, including in emerging families of quantum materials with strongly coupled electronic, spin and lattice degrees of freedom.

\section*{Acknowledgments}
\vspace{-10pt}
This work is supported by the European Union’s Horizon Europe research and innovation program under the Marie Sklodowska-Curie grant agreement 101118915 (TIMES). This work is part of the project I+D+i PID2020-112507GB-I00 QUANTA-2DMAT, funded by MCIN/AEI/10.13039/501100011033, project PROMETEO/2021/082 (ENIGMA) and SEJIGENT/2021/034 (2D-MAGNONICS) funded by the Generalitat Valenciana. This study is also part of the Advanced Materials program (project SPINO2D), supported by MCIN with funding from European Union NextGenerationEU (PRTR-C17.I1) and by Generalitat Valenciana. A. M.-S. acknowledges the Ramón y Cajal program (grant RYC2018-024024-I; MINECO, Spain). A.E.-K. acknowledges the Contrato Predoctoral Ref. PRE2021-097581.
D.S. acknowledges funding from MaX "MAterials design at the eXascale” (Grant Agreement No. 101093374) co-funded by the European High Performance Computing joint Undertaking (JU) and participating countries, and from 
the Innovation Study Isolv-BSE that has received funding through the Inno4scale project, which is funded by the European High-Performance Computing Joint Undertaking (JU) (Grant Agreement No 101118139). 
K.L. and M.B. were supported by the U.S. Department of Energy, Office of Science, Office of Advanced Scientific Computing Research and Office of Basic Energy Sciences, Scientific Discovery through Advanced Computing (SciDAC) program under Award Number DE-SC0022088 for method development, by the National Science Foundation under
Grant No. OAC-2209262 for code development, and by 
AFOSR and Clarkson Aerospace under Grants No. FA9550-21-1-0460 and
FA9550-24-1-0004 for calculations on 2D materials. This research used resources of the National Energy Research Scientific Computing Center (NERSC), a U.S. Department of Energy Office of Science User Facility located at Lawrence Berkeley National Laboratory, operated under Contract No. DE-AC02-05CH11231.


\clearpage
\newpage
\bibliography{bib.bib}

\end{document}